\documentclass[letterpaper,11pt]{article}
\pdfoutput=1 % to force arxiv to pdflatex

\usepackage{jheppub}
\usepackage{graphicx}
\usepackage{hyperref}
\usepackage{caption,subcaption} % Subfigures
\usepackage[section]{placeins}
\usepackage{amsfonts}

\newcommand{\comment}[1]{}
\newcommand{\beq}[1]{\begin{equation}\label{#1}}
\newcommand{\eeq}{\end{equation}}
\renewcommand{\O}{\mathcal{O}}
\newcommand{\C}{\mathcal{C}}

\newcommand{\del}{\partial}
\newcommand{\Del}{\nabla}
\newcommand{\M}{\mathcal{M}}
\newcommand{\bea}{\begin{eqnarray}}
\newcommand{\eea}{\end{eqnarray}}
\renewcommand{\t}{\tilde}

\title{Complexity, scaling, and a phase transition}

\author[a,b,c]{Jiayue Yang}
\author[d]{and Andrew R.~Frey}

\affiliation[a]{Center for Theoretical Physics and College of Physics, 
Jilin University\\ 2699 Qianjin St, 
Changchun 130012, People’s Republic of China}

\affiliation[b]{Department of Physics and Astronomy, University of Waterloo\\ 
220 University Ave W, Waterloo, ON N2L 3G1, Canada}

\affiliation[c]{Perimeter Institute For Theoretical Physics\\ 
31 Caroline St N, Waterloo, ON N2L 2Y5, Canada}

\affiliation[d]{Department of Physics and Winnipeg Institute for
Theoretical Physics, University of Winnipeg\\
515 Portage Avenue, Winnipeg, Manitoba R3B 2E9, Canada}

\emailAdd{j43yang@uwaterloo.ca}
\emailAdd{a.frey@uwinnipeg.ca}

\abstract{We investigate the holographic complexity of CFTs compactified
on a circle with a Wilson line, dual to magnetized solitons in AdS$_4$ and
AdS$_5$. These theories have a confinement-deconfinement phase transition
as a function of the Wilson line, and the complexity of formation acts as
an order parameter for this transition. Through explicit calculation, 
we show that proposed complexity functionals based on volume and action
obey a scaling relation with radius of the circle and further prove that
a broad family of potential complexity functionals obeys this scaling behavior.
As a result, we conjecture that the scaling law applies to
the complexity of conformal field theories on a circle in more general
circumstances.}


\begin{document}
\maketitle

\section{Introduction}

Holographic complexity is the gravity-side dual of the (circuit) complexity
of a gauge theory in the AdS/CFT correspondence, which is itself a measure of 
the Hilbert space distance of the gauge theory's state from some reference 
state. There is by now a large literature studying holographic complexity
(see \cite{arxiv:1403.5695,arXiv:1406.2678,arXiv:1509.07876,arXiv:1512.04993}
for some foundational work and \cite{arXiv:2110.14672} for more references
and a recent review); because there are continuous families of 
distance measures and possible reference states, it is clear that there should
be many formulations of holographic complexity. Indeed,
\cite{arXiv:2111.02429,arXiv:2210.09647} demonstrated that a large family
of functionals in asymptotically AdS spacetimes have the expected behavior
of complexity on thermal states. 

As a result, it is the change of complexity in parameter space rather than
the value that is physically important; often, as in black hole spacetimes,
the time derivative of complexity is the quantity of interest. It is also
important to understand the variation of complexity with other parameters,
even in static situations. To that end, we examine the holographic complexity 
of a CFT on a circle as a function of the radius and the Wilson line 
around the circle (of a $U(1)$ gauge field). The gravitational dual of the
compactified CFT with Wilson line is a magnetized generalization of the
AdS soliton first discussed by \cite{hep-th/9808079,hep-th/9803131}; 
as in the standard AdS soliton, the periodic direction of the magnetized 
solitons shrinks at a finite AdS radius 
\cite{arXiv:1205.6998,arXiv:1807.07199,arXiv:2009.14771,arXiv:2104.14572}, 
meaning that the spacetime has no horizon. Moreover, if the Wilson line
is below a critical value, the soliton has negative energy (compared to
AdS spacetime), so it is the ground state with those boundary conditions,
and the gauge theory exhibits confinement. At larger Wilson lines, the
soliton energy becomes positive, so the system has a (zero-temperature) 
confinement-deconfinement phase transition --- empty AdS with a periodic
boundary direction and a Wilson line is the ground state. 

Reynolds and Ross \cite{arXiv:1712.03732} evaluated the holographic 
complexity of the standard AdS soliton.
Here, we extend their work by evaluating 
the complexity of formation of the magnetized solitons in AdS$_4$ and AdS$_5$,
the difference in the complexity between the soliton and AdS with periodic
boundary \cite{arXiv:1509.07876,arXiv:1512.04993,arXiv:1610.08063}.
Since the ground state with large Wilson line
corresponds to periodic AdS, the complexity of formation vanishes in 
the deconfined phase and acts as an order parameter for the phase transition.
To our knowledge, this is the first investigation of complexity as an
order parameter in a first-principles confining scenario in 
holography.\footnote{though see \cite{Ghodrati:2018hss,arXiv:1808.08719} 
for studies of 
complexity in phenomenological models of confinement}
As emphasized in \cite{arXiv:2104.14572}, the state with critical Wilson line
is also supersymmetric, so tuning the Wilson line also allows us to 
examine the effect of supersymmetry breaking on complexity.

After reviewing the magnetized soliton backgrounds in section \ref{s:solitons},
we will consider the complexity of formation as calculated both as a 
volume (of a maximal spatial slice and of a spacetime region) in section
\ref{s:volume} and as an action in section \ref{s:action}.
In addition to finding the dependence of complexity on the Wilson line,
we demonstrate that the density of complexity of formation (per unit volume of
the boundary CFT) scales as the inverse $(d-1)$st power of the circumference
of the boundary circle.\footnote{Due to divergences in the UV, we must
calculate complexity with a UV cutoff. Since the divergence structure is
the same for AdS, the complexity of formation is finite; this scaling
emerges as we take the cutoff to infinity.} 
This scaling behavior persists for a large family
of possible formulations for holographic complexity, as we discuss in 
section \ref{s:scaling}.
Reynolds and Ross \cite{arXiv:1712.03732} first found this scaling for
the standard soliton (and a consistent decrease in complexity with increasing
radius for a lattice fermion model); our major results on the scaling
are that it factors from the dependence on the Wilson line and a proof that
it is universal for any holographic complexity functional obeying a few
assumptions.

We summarize our results and make some conjectures in section \ref{s:discuss}.

\section{Magnetized AdS solitons and phase transition}\label{s:solitons}

Here, we review the geometry of magnetized AdS$_{d+1}$ solitons along with the
corresponding physics of the dual gauge theory, largely following
the results of \cite{arXiv:2104.14572}. Suppose that the gauge theory
is on a flat spacetime
with one spatial coordinate $\phi$ periodically identified with period
$\Delta\phi$. In this case, the gauge theory can have a $U(1)$ 
Wilson line around
the $\phi$ direction, and there are three regular solutions of the
holographic dual gravity theory with a $U(1)$ gauge field. (This is a
bosonic subsector of gauged supergravity with 8 supercharges in either
4 or 5 dimensions, and we limit our calculations to $d=3,4$.)

The first solution is AdS with a periodic direction and a Wilson line in the
bulk. For reference, the metric and gauge field are
\beq{periodicAdS}
ds^2=\frac{r^2}{l^2}\left(-dt^2+d\vec{x}^2+d\phi^2\right)+\frac{l^2}{r^2}dr^2,
\quad A=-\frac{\Phi}{\Delta\phi} d\phi\eeq
for holonomy $\Phi$ of the boundary gauge field.
The other two solutions are generalizations of the AdS soliton; both have
metric (for $d=3,4$)
\beq{metric}
ds^2=\frac{r^2}{l^2}\left(-dt^2+d\vec{x}^2+f(r)d\phi^2\right)+
\frac{l^2}{r^2f(r)}dr^2,\quad f(r)\equiv 1-\frac{\mu l^2}{r^d}-
\frac{Q^2l^2}{r^{2d-2}}\eeq
and gauge field 
\beq{gauge} A = \sqrt{7-d}\,Q\left(\frac{1}{r^{d-2}}-\frac{1}{r_0^{d-2}}\right) 
d\phi,\quad Q\equiv \frac{1}{\sqrt{7-d}}\frac{r_0^{d-2}\Phi}{\Delta\phi}.\eeq
Like the AdS soliton, the bulk spacetimes terminate at $r_0$, the largest
root of $f(r)$; for the soliton geometries to be regular at $r_0$, $\mu$
must yield
\beq{smoothness} \Delta\phi = \frac{4\pi l^2}{r_0^2f'(r_0)}\eeq
(note that $\phi$ has units of length, so $\Delta\phi$ is a circumference).
There are two solutions to $f(r_0)=0$ and (\ref{smoothness}) for both 
values of $d$ we consider:
\beq{twosolitons}
\mu = \frac{r_0^{2d-2}-Q^2l^2}{l^2r_0^{d-2}},\quad 
r_0= \frac{2\pi l^2}{d\Delta\phi}\left(1\pm \sqrt{1-\frac{\Phi^2}{\Phi_{max}^2}}
\right)
\eeq
with $\Phi_{max} = g(d)\pi l$, $g(d)=\sqrt{(46-10d)/d}$.
The shorter soliton, ie, the solution with larger $r_0$, 
goes to the AdS soliton for $Q=0$, whereas the longer soliton merges with the
periodic AdS solution in that limit. For completeness, we note that
the two magnetic solitons have field strength
\beq{fieldstrength} F = -\sqrt{7-d}(d-2)\frac{Q}{r^{d-1}}dr\wedge d\phi.\eeq
With the usual boundary conditions of fixed geometry and gauge field,
the variables of the CFT are the periodicity $\Delta\phi$ and Wilson line
$\Phi$.  

There are two methods to determine the energy density of each solution and
therefore the ground state of the CFT with given periodicity and Wilson
line. Because it will be useful later, we give the holographic renormalization
argument here.\footnote{and a calculation following Hawking and Horowitz
\cite{gr-qc/9501014} in appendix \ref{a:energy}.}
We make a Fefferman-Graham expansion \cite{fg} of the asymptotic form of 
the bulk metric as 
\beq{fgexpansion}
ds^2 =l^2\frac{dz^2}{z^2}+ \frac{l^2}{z^2}\left(\gamma^{(0)}_{ij}+\sum_{n=d}^\infty
z^n\gamma^{(n)}_{ij}\right)dx^i dx^j;
\eeq
with $z=0$ at the boundary (and $x^i=[t,\vec x,\phi]$ in our case).
Generally, the sum can include terms with $1\leq n<d$ as well as terms
proportional to $z^n\ln(z)$ for $n\geq d$ and $d$ even, although those terms
vanish for the solitons we consider. Apparently
\beq{radialrelation}
\int\frac{dz}{z} = -\int\frac{dr}{r\sqrt{f(r)}}\quad\Rightarrow\quad
\ln(z/l) = -\ln(r/l) +\frac{\mu l^2}{2dr^d}+\cdots\eeq
using a large radius expansion. Solving iteratively,
\beq{radialrelation2}
r=\frac{l^2}{z} \left(1+\frac{\mu l^2}{2d} \frac{z^d}{l^{2d}}+\cdots\right).\eeq
In holographic renormalization (see for example 
\cite{hep-th/0002230,arxiv:1211.6347}), the boundary stress tensor is
$\langle T_{ij}\rangle= dl^{d-1}\gamma^{(d)}_{ij}/16\pi G+\cdots$, 
where $G$ is the $(d+1)$-dimensional Newton constant in the AdS spacetime and
the dots are terms that vanish on the soliton solutions.
The energy density is therefore
$\langle T_{tt}\rangle =-\mu/16\pi G l^{d-1}$. The long soliton ($r_0$ 
small) has negative $\mu$ and positive energy density always, while the
short soliton has negative energy at small $\Phi$ and positive energy 
when $\Phi\geq\Phi_S\equiv 2\pi l/[\sqrt 3]$ for $d=3$ [$d=4$].
Therefore, the system undergoes a phase transition from the short soliton
at small $\Phi$ to periodic AdS for $\Phi>\Phi_S$. Since the ground
state of the theory is either periodic AdS or the short soliton, we
will henceforth always mean the short soliton when we discuss a soliton
solution.

Because the soliton solutions cap off smoothly in the infrared rather 
than having a horizon, they exhibit confinement (like the standard $Q=0$ 
AdS soliton \cite{hep-th/9808079,hep-th/9803131}). As a result, the phase
transition from tuning
the Wilson line $\Phi$ through $\Phi_S$ is a confining/deconfining transition.
(\cite{arXiv:2104.14572} also emphasized that the short soliton is
supersymmetric at $\Phi=\Phi_S$.)

As noted above, the usual Dirichlet boundary condition on the gauge field
$\delta A_\mu=0$ means that the CFT is defined in terms of fixed Wilson line
$\Phi$; in thermodynamics, this is the grand canonical ensemble.
With an additional term in the action for the Maxwell field on the 
conformal boundary of AdS (or, precisely speaking, on a cutoff surface
at fixed large radius $r$), the natural variable of the CFT is 
$Q$ \cite{hep-th/9902170,arXiv:2104.14572}.
The boundary term in the action means that the variational problem of the 
gauge field has Neumann boundary conditions ($\delta F_{\mu\nu}=0$). 
(Multiplication of the boundary action 
term by an arbitrary coefficient leads to mixed, or Robin, boundary conditions.)
In the Euclidean theory, adding the boundary term gives the theory in the
canonical ensemble. We will return to this point later.

\section{Volume complexity of formation}\label{s:volume}

The simplest proposals for holographic complexity are given in terms of a
volume on the gravity side of the correspondence. The initial proposal
for complexity, known as CV or ``complexity=volume'' 
\cite{arxiv:1403.5695,arXiv:1406.2678}, for an asymptotically AdS spacetime is
$C_V=(d-1)V/2\pi^2 Gl$, where $V$ is the volume of a maximal volume slice 
anchored at a fixed time on the boundary. While we have written the 
normalization of the complexity with the AdS scale $l$, the choice of 
length scale is ambiguous. To compare the complexity of magnetized solitons
with varying Wilson line, we should choose a fixed length scale, but
the overall normalization is unimportant, so we choose the AdS scale for
simplicity.\footnote{The remainder of the normalization constant is chosen
so the AdS-Schwarzschild black hole saturates Lloyd's bound on the 
time derivative of complexity at late times \cite{arXiv:1712.03732}.}

All the backgrounds we consider
have both time translation and time reversal symmetries, so the maximal
volume surface is a constant time surface (we will always choose to measure
complexity at boundary time $t=0$). Because the volume diverges near
the conformal boundary, we can integrate out only to a finite radius $r_m$.
For both the magnetized solitons and periodic AdS, this volume is
\beq{volume} V = V_{\vec x}\Delta\phi \int_{r_0}^{r_m} dr\left(\frac rl\right)^{d-2}
=\frac{V_{\vec x}\Delta\phi}{d-1}\frac{r_m^{d-1}-r_0^{d-1}}{l^{d-2}},
\eeq
with $r_0\to 0$ for periodic AdS, where $V_{\vec x}$ is the volume along
the $\vec x$ directions. 

To find the complexity of formation of the soliton, we should subtract the 
corresponding volume of periodic AdS from 
(\ref{volume}) \cite{arXiv:1610.08063}. As a result,
periodic AdS has by definition vanishing complexity of formation, and any
soliton appears to have negative complexity of formation. There is a 
subtlety to consider, however. 
Rather than comparing volumes using the same cut-off 
radius $r_m$, we should compare them using the same Fefferman-Graham
coordinate $z$ as defined in (\ref{radialrelation}). If $r'_m$ is the cut off
in periodic AdS and $r_m$ in the soliton, then (\ref{radialrelation2}) 
implies that
\beq{cutoffs}r_m^{d-1} =r_m^{\prime d-1}\left( 1+ 
\frac{\mu l^2(d-1)}{2d r_m^{\prime d}}+\cdots\right). \eeq
Therefore, the divergent terms in the maximal volumes cancel as $r_m\to\infty$,
meaning the density of complexity of formation is negative:
$\C_V=-r_0^{d-1}/2\pi^2 Gl^{d-1}$ in terms of the soliton radius $r_0$.
Similarly, the difference in $r_m$ between solitons with different
values of $\Phi$ also does not contribute as we take 
$r_m\to\infty$.\footnote{As a note, suppose that we instead choose a cut off
$r'_m$ in periodic AdS such that the proper circumference of $\phi$ at the
cut off is the same in both periodic AdS and the soliton. We also see that
the difference between $r_m^{\prime d-1}$ and $r_m^{d-1}$ vanishes at infinite
cut off.}

\begin{figure}[t]
\begin{center}\includegraphics[width=0.6\textwidth]{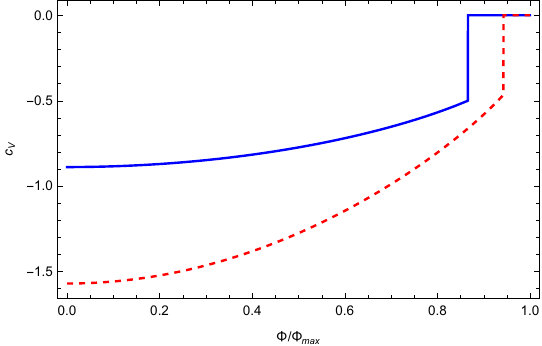}\end{center}
\caption{CV complexity of formation per unit volume for $d=3$ (solid blue)
and $d=4$ (dashed red). Plotted curves are 
$c_V\equiv G\C_V (\Delta\phi/l)^{d-1}$, so complexity scales with an inverse
power of $\Delta\phi$.\label{f:cv}}
\end{figure}

Two features of $\C_V$ are immediately apparent. First, it scales as an inverse
power of the circumference $\C_V\propto\Delta\phi^{-(d-1)}$, 
and it is discontinuous
at the phase transition since $r_0\neq 0$ at $\Phi=\Phi_S$.\footnote{See
section \ref{s:action} for the case that the natural CFT variable is $Q$
rather than $\Phi$, however. Note also that the total complexity of formation
scales as $C_V\propto \Delta\phi^{-(d-2)}$.} We show the dimensionless quantity
$c_V\equiv G\C_V (\Delta\phi/l)^{d-1}$ as a function of $\Phi/\Phi_{max}$
in figure \ref{f:cv}. Note that $c_V$ jumps from a negative value to
zero at $\Phi=\Phi_S$, the supersymmetric value, because of the deconfining
phase transition at that Wilson line. While the value of the complexity
appears similar at $\Phi_S$ for the two dimensionalities, it is not equal.

The negative value of $\C_V$ is noteworthy; \cite{arXiv:2109.06883} showed
that maximal volume slices in a wide variety of asymptotically AdS spacetimes
are always larger than maximal slices in AdS. However, they assume maximal
symmetry of the spatial slice of the conformal boundary, which the solitons
(and periodic AdS) violate, so their theorems do not apply.

Another proposal for holographic complexity, known as CV2.0, is that
$C_2\equiv V_{WDW}/Gl^2$, where $V_{WDW}$ is the spacetime
volume of the Wheeler--DeWitt (WDW) patch \cite{arXiv:1610.02038}. 
The WDW patch is the spacetime
region bounded by future- and past-directed lightsheets emitted from
the boundary time slice where the complexity is measured; once again,
we choose the length scale in the normalization as the AdS length $l$ for
simplicity and note that the overall normalization of complexity is
unimportant physically. In defining the WDW patch, we choose the boundary
conditions that both lightsheets satisfy $t=0$ at the radial cutoff $r=r_m$
as a regulator, rather than taking $t=0$ as $r\to\infty$ and cutting off
the patch at $r=r_m$. See \cite{Akhavan:2019zax,Omidi:2020oit} for a 
demonstration that these regulators are equivalent when properly defined
in the context of action complexity.

\begin{figure}[t]
\begin{center}\includegraphics[width=0.6\textwidth]{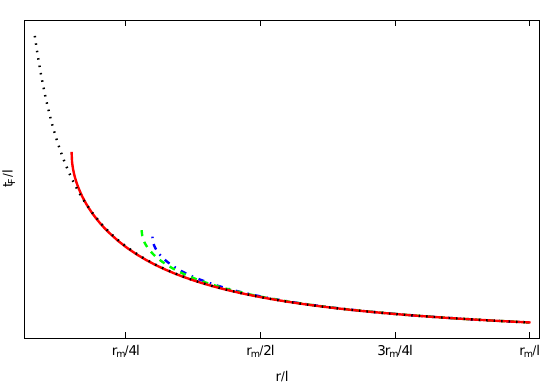}\end{center}
\caption{The future lightsheets $t_F(r)$ for
$\Phi/\Phi_{max}=$0 (dot-dashed blue), 0.5 (dashed green), 1 (solid red). 
In this instance, $\Delta\phi\, r_m/\ell^2=10\pi/3$ and $d=4$. For reference,
$t_F(r)$ for periodic AdS is dotted black.
\label{f:lightsheets}}
\end{figure}

The lightsheets on the WDW patch boundary are given by $t_F(r)$ and 
$t_P(r)=-t_F(r)$, where $F,P$ respectively designate the future and past 
lightsheets; they are given by $dt_F/dr=- l^2/r^2\sqrt{f(r)}$
and extended along $\vec x$ and $\phi$. For fixed $\Delta\phi\, r_m$,
increasing $\Phi$ lengthens the soliton (decreasing $r_0$ for the short
soliton). In addition, $t_F(r_0)$ is finite but develops a cusp
(infinite derivative). See figure \ref{f:lightsheets} for a comparison of 
$t_F$ for several values of $\Phi/\Phi_{max}$ in $d=4$ 
(we choose a small value of $\Delta\phi\, r_m$ to
emphasize the difference in the lightsheets near the soliton cap).
Note that the end of each curve at the left is the end of the spacetime at
$r_0$. For reference, we also show $t_F(r)=l^2/r-l^2/r_m$ for AdS spacetime.

\begin{figure}[t]
\begin{center}\includegraphics[width=0.6\textwidth]{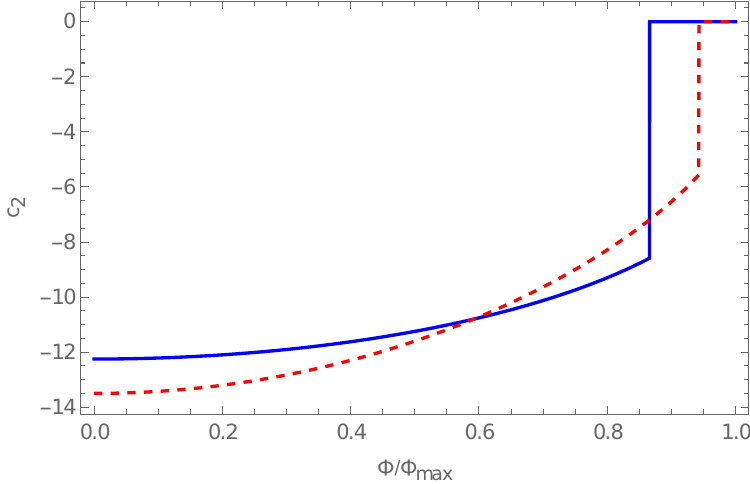}\end{center}
\caption{CV2.0 complexity of formation per boundary volume for $d=3$ 
(solid blue) and $d=4$ (dashed red). Plotted curves are 
$c_2\equiv G\C_2 (\Delta\phi/l)^{d-1}$, so complexity scales with an inverse
power of $\Delta\phi$.\label{f:cv2}}
\end{figure}

Because the WDW patch volume is
\beq{wdwvol} V_{WDW} = \int_{WDW} d^{d+1}x\sqrt{-g} =2V_{\vec x}\Delta\phi
\int_{r_0}^{r_m} dr\,\left(\frac rl\right)^{d-1} t_F(r) ,\eeq
we expect the complexity of formation to be negative --- the increase in
$t_F$ near $r_0$ is not enough to compensate for the shortening of spacetime.
In appendix \ref{a:cv2}, we describe the numerical calculation of 
$\C_2$, the density of complexity of formation, holding the
Wilson line $\Phi$ (rather than $Q$) fixed. 
Direct calculation shows that $\C_2\propto\Delta\phi^{-(d-1)}$ as we remove the 
cutoff $r_m\to\infty$, which is notably the same scaling as $\C_V$. 
We therefore
show $c_2\equiv G\C_2 (\Delta\phi/l)^{d-1}$ in figure \ref{f:cv2}. Like the CV
complexity of formation, the magnitude of $c_2$ decreases as $\Phi$ increases
and jumps to zero at the deconfining phase transition.

\section{Action complexity of formation}\label{s:action}

The action complexity (``complexity=action'' or CA)
is given by $C_A=S_{WDW}/\pi$, where $S_{WDW}$ is
the action evaluated on the WDW patch, including appropriate terms on the 
boundary of the 
patch \cite{arXiv:1509.07876,arXiv:1512.04993,arXiv:1609.00207}. 
Altogether, this action is $S_{WDW}=S_{bulk}+S_{bdy}+S_{joint}$, where
\bea
S_{bulk}&=& \frac{1}{16\pi G}
\int_{WDW} d^{d+1}x \sqrt{-g} \left(R+\frac{d(d-1)}{l^2}-\frac 14
F_{\mu\nu}F^{\mu\nu}\right) ,
\label{Sbulk}\\
S_{bdy}&=&-\frac{1}{8\pi G}
\int_F d\lambda\, d^{d-2}\vec x\, d\phi \sqrt{\gamma}\left(\kappa_F +
\Theta_F\ln|\beta l\Theta_F|\vphantom{\frac 12}\right)\nonumber\\
&&+\frac{1}{8\pi G}
\int_P d\lambda\, d^{d-2}\vec x\, d\phi \sqrt{\gamma}\left(\kappa_P+
\Theta_P\ln|\beta l\Theta_P|\vphantom{\frac 12}\right) ,\label{Slightsheet}\\
S_{joint}&=&-\frac{1}{8\pi G}\int_{F\cap P} d^{d-2}\vec x\, d\phi \sqrt{\gamma} 
\ln\left|k_F\cdot k_P/2\right| .\label{Sjoint}
\eea
The action diverges, so we regulate by integrating for $r\leq r_m$ only
and subtract the corresponding action of periodic AdS to cut-off radius
$r'_m$ related to $r_m$ by (\ref{cutoffs}); note that
the Wilson line does not contribute to the periodic AdS action because
the field strength vanishes.
The somewhat novel boundary and joint terms are by now thoroughly discussed
in the literature, so we relegate a detailed description to appendix
\ref{a:action}. We note that $\beta$ is an arbitrary parameter that in 
principle could affect the complexity; however, it cancels in the complexity
of formation when subtracting the periodic AdS action in the $r_m\to\infty$
limit. (In $d=4$, there is also a Chern-Simons term for the gauge field
in the bulk action, but it vanishes for the soliton solutions.)

Like the boundary terms (\ref{Slightsheet}) for gravitational degrees of 
freedom, the WDW patch action includes a boundary term for the Maxwell
field
\beq{Smaxbdy}
\Delta S_{bdy} = \frac{\nu}{16\pi G} \int_F d\lambda\, d^{d-2}\vec x\, 
d\phi \sqrt{\gamma} k_F^\mu A^\nu F_{\mu\nu}+
\frac{\nu}{16\pi G} \int_P d\lambda\, d^{d-2}\vec x\, 
d\phi \sqrt{\gamma} k_F^\mu A^\nu F_{\mu\nu} 
\eeq
with $\nu$ an arbitrary constant \cite{arXiv:1901.00014}.
This is the lightlike analog of the term on the AdS conformal boundary that
changes the natural variable of the CFT. However, as especially emphasized
by the ``complexity$=$anything'' program 
\cite{arXiv:2111.02429,arXiv:2210.09647}, there is no logical connection 
between $\nu$ and the boundary conditions at $r=r_m$. In other words,
$\nu$ can take any value regardless of whether $\Phi$ or $Q$ is the 
natural variable of the CFT. Also, since the complexity is given by the
action evaluated on shell, $\Delta S_{bdy}$ is equivalent to a multiple of the
bulk Maxwell action (in the absence of sources) \cite{arXiv:1901.00014}.
As a result, we can take the prefactor of $F_{\mu\nu}F^{\mu\nu}$ to
$(2\nu-1)/4$ rather than $-1/4$ in (\ref{Sbulk}) rather than adding an
additional boundary term.

The Wilson line in the CFT has several effects on the action 
complexity (evaluated at fixed periodicity $\Delta\phi$). 
As noted previously, increasing $\Phi$ lengthens the soliton (decreasing $r_0$),
tending to increase the magnitudes of $S_{bulk}$ and $S_{bdy}$.
The change in the shape of $t_F(r)$ additionally affects the
expansion $\Theta_{F,P}$ of the lightsheets at the WDW patch boundary,
primarily in the interior region.
The field strength itself gives contributes to the
Lagrangian density (including a positive semidefinite contribution 
to the curvature for $d\geq 3$); this contribution is negative for small 
$\nu$ but positive for $\nu>1/(d-1)$.

\begin{figure}[t]
\centering
\begin{subfigure}[t]{0.47\textwidth}
\includegraphics[width=\textwidth]{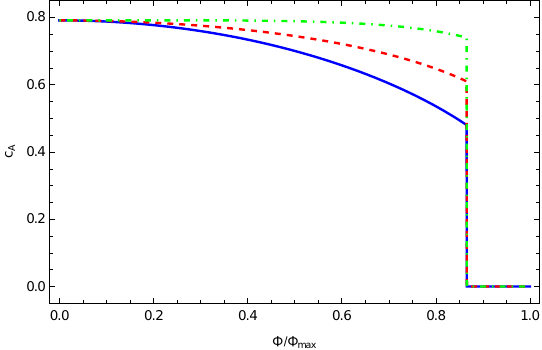}
\caption{$d=3$}\end{subfigure}
\begin{subfigure}[t]{0.47\textwidth}
\includegraphics[width=\textwidth]{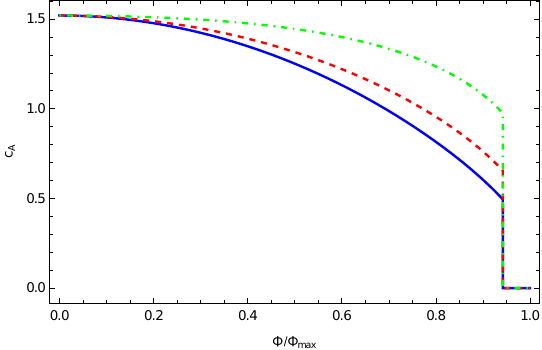}
\caption{$d=4$}\end{subfigure}
\caption{CA complexity of formation per unit boundary 
volume for $d=3,4$ as labeled
with $\nu=0$ (solid blue), $\nu=1/(d-1)$ (dashed red), and $\nu=1$
(dot-dashed green). Plotted curves are
$c_A=G\C_A(\Delta\phi/l)^{d-1}$ as a function of the Wilson line $\Phi$.
\label{f:ca}}
\end{figure}

Using the grand canonical ensemble variables $\Delta\phi,\Phi$ appropriate
to standard boundary conditions on AdS, the density of complexity of formation
$\C_A$ scales as $\Delta\phi^{-(d-1)}$, like the volume
complexities $\C_V,\C_2$, as we show in appendix
\ref{a:action}. Given that $\C_A$ is considerably more
intricate to calculate than either volume complexity, it may be surprising
that it obeys the same simple scaling relation.
Note that this scaling does not hold in the canonical ensemble
in which the complexity is evaluated as a function of $Q$. 
In that case, the scaling
property of the action implies that $\C_A\propto\Delta\phi^{-(d-1)}$
along lines of constant $Q\Delta\phi^{d-1}$ (see equations 
(\ref{gauge},\ref{twosolitons})). 

We show $c_A\equiv G\C_A(\Delta\phi/l)^{d-1}$ as a function of $\Phi$
in figure \ref{f:ca}. In contrast to $\C_V$ and $\C_2$, 
$\C_A$ is positive semi-definite
(equal zero in the deconfined phase). In the absence of the lightsheet
boundary term for the Maxwell field ($\nu=0$), the complexity is maximized
with no Wilson line and decreases toward $\Phi=\Phi_S$ (solid blue curve). 
For $\nu=1/(d-1)$, the $F_{\mu\nu}F^{\mu\nu}$ term
cancels, so the corresponding (dashed red) curve shows only the geometric 
effect of the Wilson line (increased soliton length, altered WDW patch shape)
on the complexity. Finally, for $\nu=1$ (dot-dashed green curve), 
the $d=3$ 
$\C_A$ curve increases at $\Phi=0$ and reaches a maximum for $\Phi<\Phi_S$.
In this case, the (positive) contribution of the $F_{\mu\nu}F^{\mu\nu}$ term
increases with $\Phi$, but the overall scaling with $r_0^{d-1}$ which decreases
with $\Phi$ eventually dominates. For $d=4$ and $\nu=1$, $c_A$ decreases
with increasing $\Phi$.

\section{Scaling with circumference in general complexity} \label{s:scaling}

In the previous two sections, we have seen that the complexity of formation
density $\C$ for magnetized AdS solitons
scales as $1/\Delta\phi^{d-1}$ when $\C$ is written as a
function of the grand canonical variables $\Delta\phi,\Phi$ and the UV
cutoff goes to infinity for the CV, CV2.0, and CA holographic complexity 
proposals. The details of these complexity measures and some of their 
properties (such as positivity or negativity on the soliton) are sufficiently
different that it is reasonable to suspect a general principle rather than a
coincidence at work.

Specifically, \cite{arXiv:2210.09647} argued that a large family of 
functionals of the form
\beq{any1}
C = \frac{1}{Gl^2}\int_\M d^{d+1}x\sqrt{-g}\, \mathcal{G} 
+\frac{1}{Gl}\int_{\M^+} d^d\sigma\sqrt{h}\,\mathcal{F}^+ +
\frac{1}{Gl}\int_{\M^-}d^d\sigma\sqrt{h}\, \mathcal{F}^- \eeq
all satisfy basic properties expected of holographic complexity.
Here $\M$ is a bulk region in asymptotically AdS spacetime with $\M^\pm$
respectively spacelike surfaces at the future and past portions of the 
boundary of $\M$ ($\sigma$ are worldvolume coordinates on $\M^\pm$) and
$\mathcal{G},\mathcal{F}^\pm$ are dimensionless scalar functions of the metric 
(and curvatures). Since the magnetized solitons contain matter, we generalize
to allow $\mathcal{G},\mathcal{F}^\pm$ to be functions of the gauge potential
and field strength as well. 
$\M$ itself optimizes a possibly different functional of the same form.
The boundaries of $\M^\pm$ form a joint at the time slice on the AdS boundary 
(more precisely, at the cutoff radius)
where the CFT complexity is to be evaluated; \cite{arXiv:2210.09647}
implies that additional contributions to $C$ on this joint are included,
possibly in a manner that cancels boundary terms from the $\M^\pm$ integrals
(see appendix \ref{a:action} for example).
Further, there are limits of the optimization procedure choosing $\M$ that
leads to lightlike $\M^\pm$, so the WDW patch is a possible region $\M$.

We will argue here that a general complexity
functional as in (\ref{any1}) leads to a density for complexity of formation
$\C\propto r_0^{d-1}\propto \Delta\phi^{-(d-1)}$ for the magnetized soliton
backgrounds described in section \ref{s:solitons}, when the soliton is 
considered a function of the grand canonical variables $\Delta\phi,\Phi$. 

To make this argument, we assume that the complexity of empty periodic AdS
is $C=(a V_{\vec x}\Delta\phi/G) (r_m/l)^{d-1}$ when evaluated with a cutoff 
radius $r_m$, where $a$ is a numerical constant. This is the behavior
seen in the CV, CV2.0, and CA proposals 
and is the strongest divergence possible for
the bulk term in AdS since curvature invariants are constant and the 
embedding $t_\pm(r)$ of $\M^\pm$ satisfies
$|\del_{r }t_\pm(r)|\leq l^2/r^2$ for spacelike and lightlike surfaces.
The lack of
subleading divergences (or a leading divergence of a lower power) means that
we can write this term as 
\beq{assumption}
C=\frac{aV_{\vec x}\Delta\phi}{G}\left[\left(\frac{r_0}{l}\right)^{d-1}
+(d-1) \int_{r_0}^{r_m} dr\, r^{d-2} \right] ,\eeq
which is suitable for subtraction from the soliton's complexity.

To proceed, we rewrite the metric (\ref{metric}) and gauge field
(\ref{gauge},\ref{fieldstrength}) in terms of rescaled coordinates
$\t r=r/r_0$ and $\t t=r_0 t/l$, which \cite{arXiv:1712.03732} 
introduced:\footnote{We use these coordinates to 
rewrite integrals in dimensionless form in appendices 
\ref{a:cv2},\ref{a:action}.}
\bea
ds^2&=&\t r^2\left(-d\t t^2+\frac{r_0^2}{l^2}d\vec{x}^2+
\frac{r_0^2}{l^2}f(\t r)d\phi^2\right)+
\frac{l^2}{\t r^2f(\t r)}d\t r^2,\quad f(\t r)= 1-\frac{1-\t Q^2}{\t r^d}-
\frac{\t Q^2}{\t r^{2d-2}},\nonumber\\
A&=& \sqrt{7-d}\,r_0\t Q \left(\frac{1}{\t r^{d-2}}-1\right)d\phi ,\quad
F=-\sqrt{7-d}(d-2)\,\frac{r_0\t Q}{\t r^{d-1}} d\t r\wedge d\phi ,\nonumber\\
\t Q^2 &\equiv& \frac{d^2/4}{7-d}g(d)^2\left(\frac{\Phi}{\Phi_{max}}\right)^2
\left(1+\sqrt{1-\Phi^2/\Phi_{max}^2}\right)^{-2} .\label{rescaled}
\eea
Because we are comparing the density of complexity of formation (ie,
complexity per volume $V_{\vec x}\Delta\phi$) across different values of 
$\Delta\phi$ and $\Phi$, we cannot rescale the $\vec x$ and $\phi$ coordinates.
Likewise, the UV cutoff is at a fixed value of the un-rescaled
radial coordinate $r=r_m$ corresponding to fixed Fefferman-Graham coordinate.

In these coordinates, the tensor components $g_{\mu\nu},A_\mu,F_{\mu\nu}$ 
contain precisely one factor of $r_0$ for each leg along $\vec x$ or $\phi$.
It is straightforward to check that the same is true of the Riemann tensor
with all indices lowered. The Christoffel symbols are also such that covariant
derivatives in the $\vec x,\phi$ directions add a factor of $r_0$ but
those in the $\t t,\t r$ directions do not (this argument also relies on
translation invariance in the $\vec x,\phi$ directions). As a result, any
bulk scalar function $\mathcal{G}$ constructed from the metric, the Riemann
tensor, the gauge field, or their covariant derivatives is independent of
$r_0$ --- the inverse metric components needed for contractions cancels them.
In addition, on either spatial surface $\M^\pm$, the normal vector $n_\mu$ 
can have components only in the $\t t,\t r$ directions due to translational
invariance, and its only nonzero covariant derivatives have only those legs
also. As a result, the extrinsic curvature has no factors of $r_0$. 
Similar arguments apply to the curvature $\kappa$ and expansion $\Theta$
if $\M^\pm$ is lightlike. Therefore, assuming $\mathcal{F}^\pm$ are constructed
only from the metric, gauge field, and those curvatures, they contain
no explicit factors of $r_0$ either. Altogether, any functional $C$ of the
form (\ref{any1}) is proportional to $r_0^{d-1}$, with all those factors
arising from determinants of the metric, except for a dependence on the
region of integration.

Under these assumptions, the density of complexity of formation takes the
form 
\beq{any2}
\C = \frac{1}{G} \left(\frac{r_0}{l}\right)^{d-1} \left\{ -a +
\int_1^{r_m/r_0} d\t r \left[ \left(\t t_+(\t r)-\t t_-(\t r)\right)
\mathfrak{g}(\t r)+\mathfrak{f}^+(\t r;\t t_+(\t r))+
\mathfrak{f}^-(\t r;\t t_-(\t r))\right]\right\} ,
\eeq
where $\mathfrak{g},\mathfrak{f}^\pm$ are the result of evaluating 
$\mathcal{G},\mathcal{F}^\pm$ on the soliton (with the AdS complexity 
subtracted), and $\t t_\pm(\t r)$ describe the surfaces $\M^\pm$. 
The dependence of $\mathfrak{f}^\pm$ on $\t t_\pm(\t r)$ includes dependence on
the derivatives of that embedding function.
Here we give the case that $\M^\pm$ are spacelike, but the
timelike case is similar (see appendix \ref{a:action} for the example
of CA complexity).
From the above discussion, $\mathfrak{g},\mathfrak{f}^\pm$ must be independent
of $r_0$, so the only possible dependence on $\Delta\phi$ is through the
upper limit of integration or implicitly in $\t t_\pm$. However, $\t t_\pm$
are given by optimization of a functional of the form (\ref{any1}), in which
dependence on $r_0$ scales out of the integrand. Therefore, $\t t_\pm$
depends on $r_0$ only through the boundary condition $\t t_\pm(\t r=r_m/r_0)=0$.
However, $\C$ is by design convergent in the $r_m\to\infty$ limit, so 
the integral cannot depend on $r_0$ in that limit. The overall scaling
is therefore $\C\propto r_0^{d-1}\propto \Delta\phi^{-(d-1)}$.

\section{Discussion and conjecture}\label{s:discuss}

To summarize, we have evaluated several proposals for holographic complexity
for a 3- or 4-dimensional CFT on a circle (times Minkowski spacetime) 
and a $U(1)$ gauge field,
which corresponds to magnetized AdS soliton backgrounds through gauge/gravity
duality. One immediate observation is that ``complexity=volume'' proposals
(both CV and CV2.0) have negative complexity of formation, while the 
complexity of formation for the ``complexity=action'' proposal is positive.
As a result, we speculate that the reference states for the CV and CV2.0
proposals lie in a different region of the CFT's Hilbert space than the
reference states for CA complexity proposals (which may vary 
with varying boundary conditions for the gauge field). Another interpretation
is that the volume and action complexity proposals have similar reference
states but very different distance measures, though that seems less likely 
to lead to the observed sign change.

As emphasized by \cite{arXiv:2104.14572}, the compactified CFT 
exhibits two behaviors near the critical Wilson line 
value $\Phi=\Phi_S$: supersymmetry breaking for $\Phi\neq\Phi_S$ and a
transition from the magnetized soliton (confining) phase to the periodic
AdS (deconfined) phase. The latter effect seems to be more important 
for the complexity; the complexity of formation acts like an order parameter,
vanishing in the deconfined phase and changing discontinuously at the 
transition. This is perhaps unsurprising given the nature of the phase 
transition with periodic AdS as the ground state in the deconfined phase.
On the other hand, the degree of supersymmetry breaking does not seem to 
influence the complexity; $\C$ is constant for $\Phi>\Phi_S$ (by definition)
but $|\C|$ grows generically as $\Phi$ decreases below $\Phi_S$. (With
further decreases in $\Phi$, $|\C|$ decreases again in some cases.)

A striking result is that the dependence of the complexity of formation
on the grand canonical variables $\Delta\phi,\Phi$ factorizes, with
the density behaving as $\C=h(\Phi)/\Delta\phi^{d-1}$, extending the 
original observation of \cite{arXiv:1712.03732} for the soliton without
magnetic field to magnetized solitons (in $d=3,4$ only, however).
This scaling law holds for the three specific forms of 
holographic complexity that we calculated; we further showed that the
general holographic complexity functional as advocated by the
``complexity=anything'' program \cite{arXiv:2111.02429,arXiv:2210.09647}
has the same scaling, under some mild assumptions. Therefore, we make
a series of progressively stronger conjectures; the first is simply that this
scaling extends to all magnetic AdS solitons of any dimensionality $d$.
Next, we conjecture that the density of 
complexity of formation for any $d$-dimensional
holographic field theory on a circle scales as the inverse $(d-1)$st power
of the circumference, when the theory is written in terms of grand
canonical variables. 
A stronger conjecture is that this scaling is true
for the complexity of formation for any conformal field theory on a circle
(or possibly even any quantum field theory).

A critical point in our arguments is the finiteness of the complexity of 
formation, which allows us to remove the UV cutoff (ie, take $r_m\to\infty$).
This is manifestly true since the soliton is asymptotically AdS, so the 
complexity of the soliton has the same divergence structure as periodic AdS.
However, for the complexity of formation to scale with the circumference
as conjectured, we assumed that the complexity of AdS itself must be a pure
divergence $\propto r_m^{d-1}$, the strongest possible divergence. If we
presuppose that the density of complexity of formation should scale as 
$\Delta\phi^{-(d-1)}$, this gives us instead an additional requirement
for a functional to describe holographic complexity (beyond the switchback
effect and linear growth at late time in black hole backgrounds).  
Since this seems like an unnatural condition for theories with a scale,
it may indicate that complexity may not have a simple scaling behavior in
non-conformal theories.

Next, we note that the Casimir energy density of a $d$-dimensional 
field theory on one finite dimension scales as $d$ powers of the inverse 
length of that dimension. In fact, we can see this behavior in the holographic
energy density as reviewed in section \ref{s:solitons}.
It is intriguing to consider whether there is a deeper connection between the 
scaling of the Casimir energy and of complexity of formation. On the other hand,
both scaling laws may simply arise because both are given by integration over
field modes in the noncompact directions of the field theory.

Finally, it is worth considering the connection of our results to those of
\cite{Andrews:2019hvq} regarding the complexity of another type of soliton
in AdS$_5$. There are two major distinctions from our work: first, the solitons
considered in \cite{Andrews:2019hvq} have positive energy so are never the
ground state; second, they are asymptotic to AdS$_5$ in global coordinates,
so the compact boundary dimension is instead one of the angular directions
of the boundary $S^3$ and cannot have arbitrary circumference.
As a result, it is difficult to make a direct comparison with our work.
Nonetheless, \cite{Andrews:2019hvq} observe that the complexity of formation
for those solitons also obeys a scaling law, in their case with the 
thermodynamic volume of the soliton. Since \cite{AlBalushi:2020rqe} observes
a similar scaling for black holes, it would be interesting in the future
to determine what kind of relationship there may be between that scaling
law and the one we have discussed here.

\acknowledgments
JY would like to thank Robert Mann, Niayesh Afshordi, Haijun Wang,
and Wencong Gan for encouragement in
pursuing this project. The work of ARF was supported by the
Natural Sciences and Engineering Research Council of Canada Discovery Grant
program, grants 2020-00054. The work of JY was supported by the Mitacs 
Globalink program.

\appendix
\section{Energy density from curvature}\label{a:energy}
An alternative method to determine the energy density of asymptotically AdS
backgrounds is to compare the extrinsic curvature of a boundary surface at
a cutoff radius $r_{m}$ to that of one with identical geometry in pure
AdS, as in \cite{gr-qc/9501014}.

Specifically, we choose a surface at fixed $t$ and large $r$. The energy of
the spacetime is
\beq{energyK}
E = -\frac{1}{8\pi G}\int N(K-K_0)\eeq
with the integral over the surface. Here, $N$ is the lapse, $K$ is the 
extrinsic curvature of the surface in the spacetime in question (the
solitons in our case), and $K_0$ the curvature of the surface in empty AdS.
The curvature is given by $K = n^\mu\del_\mu A$, where $n^\mu$ is the radial unit
vector and $A$ is the area of the fixed $t,r$ surface. 

For the soliton backgrounds with a surface at $r=r_m$, 
the lapse is $N=r_m/l$, the radial unit vector has
only one nontrivial component $n^r=r_m\sqrt{f(r_m)}/l$, and the area of the
surface is (in the $r_m\to\infty$ limit)
\beq{surfacearea} A =V_{\vec x}\Delta\phi\left(\frac{r_m}{l}\right)^{d-1} 
\sqrt{f(r_m)} \quad\Rightarrow\quad \int NK = 
\frac{V_{\vec x}\Delta\phi}{l^{d+1}}\left[(d-1)r_m^d+\left(1-\frac d2\right)
\mu l^2\right] ,\eeq
with $V_{\vec x}$ as in the complexity. $NK_0$ is the same with $\mu=0$, except
we must take $\Delta\phi\to \sqrt{f(r_m)}\Delta\phi$ (changing the asymptotic
periodicity), so the two surfaces have the same proper size at the cutoff
radius $r_m$, as emphasized in \cite{hep-th/9808079}. In the end, we find
$E=-V_{\vec x}\Delta\phi \mu/16\pi G l^{d-1}$, consistent with the holographic
stress tensor.

Rather than match the proper size of the cutoff surfaces in
the solition and periodic AdS spacetimes, we could instead take cutoff
surfaces at the same Fefferman-Graham coordinate $z$, as in the complexity
of formation. Amusingly, this gives the same result for the energy for
the magnetized solitons, though it is not a general feature of this formalism
for the energy of asymptotically AdS spacetime.

\section{Spacetime volume calculations}\label{a:cv2}

We determine $t_F(r)=-t_P(r)$ by numerical integration with
$t=0$ at $r=r_m$; it turns out that the $Q=0$ case can be written in terms
of hypergeometric functions, but that does not generalize. It is 
straightforward to see that the $r_0 t_F/l$ vs $r/r_0$ curve falls in a 
two-parameter family depending on $\Delta\phi\, r_m$ and $\Phi/\Phi_{max}$;
the key point to note is that $f(r)$ is independent of $\Delta\phi$ when
written in terms of a dimensionless radial variable $\tilde r =r/r_0$ for
fixed $\Phi$. Then 
\beq{tFscaling} t_F(r) =\frac{l^2}{r_0} \int_{\t r}^{r_m/r_0} 
\frac{d\t r'}{\t r'^2\sqrt{f(\t r')}}\equiv \frac{l}{r_0}\t t_F(\t r) ,\eeq
where $\t t_F$ depends on the circumference only through the boundary 
condition that $\t t_F=0$ at 
$\t r=r_m/r_0\propto \Delta\phi\, r_m$. We evaluate $t_F(r)$ numerically.

To find the complexity of formation, we also need the WDW patch volume for
periodic AdS, which is
\beq{volwdwads} V_{AdS} =2V_{\vec x}\Delta\phi\int_0^{r'_m} dr
\left(\frac rl\right)^{d-1} \left(\frac{l^2}{r}-\frac{l^2}{r_m}\right)
=\frac{2V_{\vec x}\Delta\phi\, l^2}{d(d-1)}\left(\frac{r'_m}{l}\right)^{d-1} .
\eeq
As before, the difference between $r'_m$ and $r_m$ does not change $V_{AdS}$
as we allow $r_m\to\infty$ based on equation (\ref{cutoffs}). 

Since we must evaluate $V_{WDW}$ numerically, we write $V_{AdS}$ as 
\beq{volwdwads2} 
V_{AdS} =\frac{2V_{\vec x}\Delta\phi\, l}{d}\left\{ \frac{l}{d-1}\left(
\frac{r_0}{l}\right)^{d-1}+\int_{r_0}^{r_m} dr\left(\frac rl\right)^{d-2}\right\}
\eeq
and subtract integrands. The boundary density of 
the complexity of formation is
\bea \C_2&=&\frac 2G \left\{ -\frac{1}{d(d-1)}\left(\frac{r_0}{l}
\right)^{d-1}+\int_{r_0}^{r_m}dr\left(\frac rl\right)^{d-1}\left[
\frac{t_F(r)}{l^2}-\frac 1d \frac{1}{r}\right]\right\}\nonumber\\
&=& \frac 2G\left(\frac{r_0}{l}\right)^{d-1}\left\{ -\frac{1}{d(d-1)}+
\int_1^{r_m/r_0}d\t r\, \t r^{d-1}\left[\frac{\t t_F(\t r)}{l}-\frac 1d
\frac{1}{\t r}\right]\right\} . \label{c2integral}
\eea
We see that the factor in braces depends on $\Delta\phi$ only through the
upper limit on the integral and the boundary condition on $\t t_F$, and it 
only appears in the combination $r_m\Delta\phi$. As a result, the quantity
in braces is independent of $\Delta\phi$ in the limit that we remove the UV
cutoff $r_m\to\infty$. Then $\C_2\propto \Delta\phi^{-(d-1)}$ due to the
overall prefactors of $r_0$.

To evaluate the numerical integral, we verified that it changes less than
1\% between values $r_m\Delta\phi=400l^2$ and $500l^2$ for $\Phi=0$,
which has the smallest ratio $r_m/r_0$ at any $\Delta\phi$. 
We therefore use $r_m\Delta\phi=500l^2$.

\section{Action calculations}\label{a:action}

Here we give details of our calculation of the action in equations
(\ref{Sbulk},\ref{Slightsheet},\ref{Sjoint}). $\lambda$ is the lightlike
parameter along each lightsheets on the future and past boundaries, with
$\lambda$ increasing into the future in each case. The vectors 
$k_F,k_P\equiv dx/d\lambda$ are the future-directed lightlike vectors along 
the corresponding lightsheets,
and $\kappa_{F,P}$ is defined by $k^\mu\Del_\mu k^\nu=\kappa k^\nu$.
$\gamma_{ij}$ is the induced metric on the spatial slices at fixed 
$\lambda$ along the lightsheet, and $\Theta_F,\Theta_P$ are the expansions of 
the lightsheets defined by the logarithmic derivative of $\sqrt\gamma$ 
with respect to $\lambda$. The $\Theta_{F,P}$ terms are counterterms needed
to ensure reparameterization invariance of $S_{bdy}$, and $\beta$ is an 
arbitrary parameter that sets the length scale of the counterterm. If we
take $\beta$ the same for both lightsheets and for all backgrounds, we will
see that $\beta$ cancels in the complexity of formation.

Again, see figure \ref{f:lightsheets} for a sample of $t_F(r)$ at 
fixed $\Delta\phi\, r_m$ and a variety of $\Phi/\Phi_{max}$.
The main difference in the shape of the lightsheets is due to the reduction in
$r_0$ as $\Phi$ increases; $t_F\to (1/r-1/r_m)$ at large radius, as in periodic
AdS. It is 
also straightforward to see that $\kappa_{F,P}=0$ if we choose $\lambda$
such that $dr/d\lambda = \mp\alpha_{F,P}\sqrt{f(r)}$ for arbitrary positive
dimensionless constants $\alpha_{F,P}$. These derivatives give the 
$\mu=t$ and $r$ components of $k_{F,P}^\mu$. The induced metric for spatial 
slices of either lightsheet gives $\sqrt\gamma=\sqrt{f(r)}(r/l)^{d-1}$.

Because $dr/d\lambda$ has opposite sign on the past and future lightsheets,
we can see that their contributions to $S_{bdy}$ are identical when written 
as an integral over radius. For notational convenience, we define
\beq{expansion}\Theta_{F,P}(r)=\frac{dr}{d\lambda} \left(\frac{d-1}{r}
+\frac{1}{2f(r)}\frac{df}{dr}\right)\equiv \frac{dr}{d\lambda} \Theta(r)\ .\eeq
The identity $\sqrt\gamma\Theta = d\sqrt\gamma/dr$ allows us to convert
several boundary terms to joint terms, and we perform an additional integration
by parts after adding and subtracting $\ln(r/l)$
to cancel a logarithmic divergence between the joint and boundary
terms. All terms including $\alpha_{F,P}$ cancel between $S_{bdy}$ and $S_{joint}$.

To simplify the bulk contribution, we use the trace of the Einstein equation,
which gives the well-known identity
\beq{einsteintrace}
R =-\frac{d(d+1)}{l^2} +\frac 14 \frac{d-3}{d-1}F_{\mu\nu}F^{\mu\nu} .
\eeq
The Maxwell term without the additional boundary term ($\nu=0$) 
is however more negative, so the field strength gives an 
overall negative contribution to the bulk action. With the Maxwell boundary
term, the terms proportional to $(F_{\mu\nu})^2$ give a positive contribution
for $\nu>1/(d-1)$.

After simplification, the total soliton action is therefore
\bea
S_{bulk}+S_{bdy}+S_{joint} &=& \frac{V_{\vec x}\Delta\phi}{8\pi G} \int_{r_0}^{r_m} dr\,
\left(\frac rl\right)^{d-1}\left\{2\sqrt{f(r)}\left[ \frac 1r+\Theta(r)
\left(\ln(r\Theta(r))+\frac 12\ln(f(r))\right)\right]\right.\nonumber\\
&&\left. -t_F(r)\left[\frac{2d}{l^2}+\left(1-\nu(d-1)\vphantom{\frac 12}
\right)\frac{(d-2)^2}{d-1}
\frac{r_0^{2d-4}\Phi^2}{\Delta\phi^2}\frac{1}{r^{2d-2}}\right]\right\}\nonumber\\
&&+\frac{V_{\vec x}\Delta\phi}{4\pi G} 
\left(\frac{r_m}{l}\right)^{d-1}\sqrt{f(r_m)}\ln(\beta) .
\eea
Note that the arbitrary constants $\alpha_{F,P}$ cancel between boundary
and joint terms.
Because $f(r)\to 1$ and $r\Theta\to d-1$ at large $r$, the divergences
are the same as empty periodic AdS. Therefore, we remove the divergences
by subtracting the action of periodic AdS with cut-off radius $r'_m$
as in (\ref{cutoffs}), which gives the complexity of formation.
Since $r_m=r'_m(1+\O(1/r_m^{\prime d}))$, we take $r'_m=r_m$; see further
discussion below. To aid in numerical cancellation of divergences, we write
the action of periodic AdS as
\beq{AdSaction}
S_{AdS} = \frac{V_{\vec x}\Delta\phi}{4\pi G}\left\{\left(\frac{r_m}{l}\right)^{d-1}
\ln(\beta)+\ln(d-1)\left[
\left(\frac{r_0}{l}\right)^{d-1}+\int_{r_0}^{r_m} dr\,\left(\frac{d-1}{l}\right)
\left(\frac{r}{l}\right)^{d-2}\right]\right\}.\eeq
The $\beta$-dependent terms cancel for $r_m\to\infty$ because
$f-1\sim\O(r_m^{-d})$, so we henceforth drop them.

A particular rescaling of the integration variable as described in 
\cite{arXiv:1712.03732} is useful both for numerical calculation of the
action and for proving scaling properties of the complexity. The function 
$f$ written in terms of $\tilde r=r/r_0$ is independent of $r_0$ and therefore
of $\Delta\phi$ (after solving for $\mu$ and $Q$ in terms of $\Delta\phi$
and $\Phi$), as discussed in appendix \ref{a:cv2}. 
Similarly, $\tilde t_F(\tilde r)\equiv r_0 t_F/l$ depends on
$\Delta\phi$ only through the boundary condition at 
$r_m/r_0\propto r_m\Delta\phi$. Then the total action is
\beq{actiontotal}
S=S_{bulk}+S_{bdy}+S_{joint}-S_{AdS} = \frac{V_{\vec x}\Delta\phi}{8\pi G}
\left(\frac{r_0}{l}\right)^{d-1} \left[-\ln(d-1)+I(r_m/r_0,\Phi)
\vphantom{\frac 12}\right],\eeq
where $I(r_m/r_0,\Phi)$ is the integral
\bea
I(r_m/r_0,\Phi) &=& \int_1^{r_m/r_0}d\t r\,\t r^{d-1} \left\{2\sqrt{f(\t r)}
\left[\frac{1}{\t r}+\t\Theta(\t r)\left(\ln(\t r\t\Theta(\t r))+\frac 12
\ln(f(\t r))\right)\right] \right.\nonumber\\
&&\left. -\frac{\t t_F(\t r)}{l}\left[2d+\left(1-\nu(d-1)\vphantom{\frac 12}
\right)\frac{d^2(d-2)^2}{d-1} \frac{g(d)^2}{4\t r^{2d-2}}
\frac{\Phi^2/\Phi_{max}^2}{\left(1+\sqrt{1-\Phi^2/\Phi_{max}^2}\right)^2}
\right]\right.\nonumber\\
&&\left.-2\ln(d-1)\frac{d-1}{\t r}\right\}\label{Sdimensionless}
\eea
and $\t\Theta(\t r) =r_0\Theta=(d-1)/\t r -(df/d\t r)/2f$ is the dimensionless
expansion. A key property to note is that, when written in terms of the
Wilson line $\Phi$, $I$ also depends on $\Delta\phi$ only through the limit
of integration $r_m/r_0\propto r_m\Delta\phi$. But the subtraction of $S_{AdS}$
means that $I$ converges as the upper limit goes to infinity --- to a 
value independent of $\Delta\phi$. As a result, when we remove the cut off, 
the density of complexity of formation is 
$\C_A\propto (\Delta\phi/l)^{-(d-1)}$ for standard boundary conditions with the
scaling coming from the prefactor of $r_0^{d-1}$.

Since the complexity of formation is the finite $r_m\to\infty$ value, we
have evaluated (\ref{Sdimensionless}) for $\Phi=0$, where $r_0$ is largest,
and checked where the integral has converged sufficiently. In both $d=3$
and $d=4$, $I(r_m/r_0,0)$ changes by less than 1\% between 
$r_m\Delta\phi=400l^2$ and $500l^2$ for $\Phi=0$ (which is the shortest
soliton). We therefore use $r_m\Delta\phi=500l^2$
for all numerical calculations of the complexity.

\bibliographystyle{JHEP}
\bibliography{solitoncomplexity}

\end{document}